\documentclass[superscriptaddress,twocolumn,prb,showpacs,preprintnumbers,amsmath,amssymb]{revtex4}

\newcommand{ \ybco }{\mbox{YBa$_2$Cu$_3$O$_6$}}
\newcommand{ \nco }{\mbox{NaCu$_2$O$_2$}}
\newcommand{ \lco }{\mbox{LiCu$_2$O$_2$}}
\newcommand{ \lcvo }{\mbox{LiCuVO$_4$}}
\newcommand{ \OCO }{\mbox{$\rm O$-$\rm Cu^{1+}$-$\rm O$}}
\newcommand{ \srxx }{\ensuremath{\sigma_{1}^{\perp}}}
\newcommand{ \sryy }{\ensuremath{\sigma_{1}^{\parallel}}}
\newcommand{\px}{\ensuremath{p_{x}}}
\newcommand{\py}{\ensuremath{p_{y}}}

\newcommand{\dxy}{\ensuremath{d_{xy}}} 
\newcommand{\dxz}{\ensuremath{d_{zx}}}
\newcommand{\dyz}{\ensuremath{d_{yz}}}

\usepackage{graphicx}
\usepackage{dcolumn}
\usepackage{bm}
\usepackage{ulem}
\usepackage[USenglish]{babel}
\usepackage{color}
\usepackage{verbatim}
\usepackage{amssymb}
\usepackage{amstext}
\usepackage{bbding}
\usepackage{pifont}

\begin{document}

\title{Anisotropic optical response of the mixed-valent Mott-Hubbard insulator NaCu$_{2}$O$_{2}$}

\author{Y.~Matiks}
\email[]{Y.Matiks@fkf.mpg.de}
\affiliation{Max Planck Institute for Solid State Research, Heisenbergstr. 1, D-70569 Stuttgart, Germany}
\author{A.~N.~Yaresko}
\affiliation{Max Planck Institute for Solid State Research, Heisenbergstr. 1, D-70569 Stuttgart, Germany}
\author{K. Myung-Whun}
\affiliation{Max Planck Institute for Solid State Research, Heisenbergstr. 1, D-70569 Stuttgart, Germany}
\affiliation{Department of Physics and IPIT, Chonbuk National University, Jeonju 561-756, Korea}
\author{A.~Maljuk}
\affiliation{Leibniz Institute for Solid State and Materials Research IFW Dresden, Helmholtzstr. 20, D-01171 Dresden, Germany}
\affiliation{Max Planck Institute for Solid State Research, Heisenbergstr. 1, D-70569 Stuttgart, Germany}
\author{P. Horsch}
\affiliation{Max Planck Institute for Solid State Research, Heisenbergstr. 1, D-70569 Stuttgart, Germany}
\author{B. Keimer}
\affiliation{Max Planck Institute for Solid State Research, Heisenbergstr. 1, D-70569 Stuttgart, Germany}
\author{A.~V.~Boris}
\email[]{A.Boris@fkf.mpg.de}
\affiliation{Max Planck Institute for Solid State Research, Heisenbergstr. 1, D-70569 Stuttgart, Germany}

\date{\today}

\begin{abstract}
We report the results of a comprehensive spectroscopic ellipsometry study of $\rm NaCu_2O_2$, a compound composed of chains of edge-sharing $\rm Cu^{2+}O_4$ plaquettes and planes of Cu\(^{1+}\) ions in a \OCO \ dumbbell configuration, in the spectral range \(0.75 - 6.5\) eV at temperatures \(7-300\) K. The  spectra of the dielectric function for light polarized parallel to the $\rm Cu^{1+}$ planes reveal a strong in-plane anisotropy of the interband excitations. Strong and sharp absorption bands peaked at 3.45 eV (3.7 eV) dominate the spectra for polarization along (perpendicular) to the $\rm Cu^{2+}O_2$ chains. They are superimposed on flat and featureless plateaux above the absorption edges at 2.25 eV (2.5 eV). Based on density-functional calculations, the anomalous absorption peaks can be assigned to transitions between bands formed by $\rm Cu^{1+}$ 3$d_{xz}$($d_{yz}$) and $\rm Cu^{2+}$ 3$d_{xy}$ orbitals, strongly hybridized with O $p$-states. The major contribution to the background response  comes from transitions between $\rm Cu^{1+}$  3$d_{z^{2}}$ and 4$p_x$($p_y$) bands. This assignment accounts for the measured in-plane anisotropy. The dielectric response along the $\rm Cu^{2+}O_2$ chains develops a weak two-peak structure centered at 2.1 and 2.65 eV upon cooling below $\sim 100$ K, along with the appearance of spin correlations along the $\rm Cu^{2+}O_2$ chains. These features bear a striking resemblance to those observed in the single-valent $\rm Cu^{2+}O_2$ chain compound $\rm LiCuVO_4$, which were identified as an exciton doublet associated with transitions to the upper Hubbard band that emerges as a consequence of the long-range Coulomb interaction between electrons on neighboring $\rm Cu^{2+}$ sites along the chains. An analysis of the spectral weights of these features yields the parameters characterizing the on-site and long-range Coulomb interactions.
\end{abstract}


\maketitle

\section{\label{sec:Intro}Introduction}

Copper oxides with quasi-one-dimensional electronic structure have drawn much attention because of their unusual magnetic properties and the variety of ground states originating from strong electronic correlations in different lattice architectures. Compounds composed of chains of edge-sharing $\rm Cu^{2+}O_4$ plaquettes have the peculiar property that the magnitude of the nearest-neighbor hopping matrix element along the chains is anomalously small due to the orthogonality of the $2p_{\sigma}$ orbitals on oxygen ions adjacent to the $\rm Cu$ ion. The interplay between short- and long-range interactions generates spiral magnetism in Mott-insulating compounds \cite{Mae01, Miz98, Mot96} and charge density waves in doped compounds.\cite{Hor05,Rai08} By virtue of their exceptionally narrow electronic bandwidths, these compounds also provide a highly favorable platform for the investigation of exciton formation and the interplay between spin and charge correlations in the cuprates. \cite{Bar02,Jec03,Geb97,Gal97,May06}

In optical experiments, Zhang-Rice singlet excitations are generated in the charge transfer process $d^9_{i\uparrow} d^9_{j\downarrow} \rightarrow (d^9L_h)_i d^{10}_j$ between $\rm Cu^{2+}O_4$ plaquettes at sites $i$ and $j$.\cite{Zha88} In the final state, the spins form singlets leaving an
oxygen-ligand hole on one of the two Cu$^{2+}$O$_4$ plaquettes. The band formed by the $d^{10}$ states on the other plaquette can be regarded as the upper Hubbard band, in formal analogy to the single-band Hubbard model. Using spectroscopic ellipsometry, Matiks {\it et al.} have recently demonstrated that Mott-Hubbard excitons are the lowest accessible states for holes in the single-valent chain cuprate $\rm LiCuVO_4$. \cite{Mat09} For photon polarization along the $\rm Cu^{2+}O_2$ chains, a weak but well-resolved two-peak structure whose spectral weight is strongly enhanced upon cooling near the magnetic ordering temperature has been identified as an exciton doublet. These results have not only persuasively demonstrated the formation of Mott-Hubbard excitons, but also quantified various important characteristic energy scales, such as the local Hubbard and long-range Coulomb interactions, the nearest-neighbor (NN) and next-nearest-neighbor (NNN) hopping parameters, and the resulting superexchange energies.

In order to explore the generality of the excitonic states observed in $\rm LiCuVO_4$, we have carried out a comprehensive ellipsometric study of the complex dielectric function of single crystals of $\rm NaCu_2O_2$, a mixed-valent Mott insulator composed of chains of edge-sharing $\rm Cu^{2+}O_4$ plaquettes and \OCO \ dumbbells.\cite{Mal04} Like $\rm LiCuVO_4$, $\rm NaCu_2O_2$ exhibits helical magnetic order at low temperatures due to competing ferromagnetic NN and antiferromagnetic NNN superexchange interactions within the $\rm Cu^{2+}O_2$ edge-sharing chains.\cite{Mal04,Cap05,Cap10,Lei10} Unlike in $\rm LiCuVO_4$, the chains in $\rm NaCu_2O_2$ are separated by nonmagnetic Cu\(^{1+}\) ions in \OCO \ dumbbell complexes along the $c$ axis. The presence of this additional structural unit in other copper oxides including $\rm YBa_2Cu_3O_6$, the parent compound of a well known family of superconductors, provides further motivation for our study.

In a recent ellipsometry study of {\lco}, a compound that is isostructural and isoelectronic to {\nco}, Pisarev {\it et al.} reported a strong and narrow absorption peak at about 3.3 eV and attributed this excitation to the \OCO \ dumbbells.\cite{Pis06} This feature obscures the weak exciton bands in the optical conductivity along the $\rm Cu^{2+}O_2$ chains that are expected following the analogy with $\rm LiCuVO_4$. Detailed information about the in-plane dielectric anisotropy is required in order to clearly separate the contribution of the $\rm Cu^{2+}$ $(d^{9})$ and $\rm Cu^{1+}$ $(d^{10})$ states to the dielectric response and to avoid ambiguity in the data interpretation. However, Li-Cu chemical inter-substitution and crystallographic twinning in {\lco} crystals obliterate the dielectric anisotropy in the \textit{ab} plane. Experiments on single crystals of {\nco} offer the chance to elucidate the intrinsic dielectric anisotropy of this class of compounds. Indeed, recent x-ray spectroscopy and neutron diffraction investigations confirmed the superior quality of {\nco} single crystals which, unlike {\lco}, are not prone to twinning and disorder.\cite{Cap05,Cap10,Lei10} The absence of crystal defects can explain the apparent absence of multiferroic behavior in {\nco}. \cite{Cap10,Lei10}

In this paper, we report a comprehensive ellipsometric study of the dielectric function anisotropy of NaCu\(_{2}\)O\(_{2}\) in the spectral range 0.75 -- 6.5 eV and its interpretation based on band-structure calculations. The paper is organized as follows. Section II describes experimental details and results.
In Section III A the discussion is focused on the behavior of the exciton bands identified in the optical conductivity along the $\rm Cu^{2+}O_2$ chains. The anomalous optical absorption is discussed in Section III B, and in Section III C local spin density approximation (LSDA) calculations are reported in order to explain the observed anomalies and the anisotropy of the optical response. Finally, our conclusions are summarized in Section IV.

\section{Experimental Details and Results}

Single crystals of NaCu$_{2}$O$_{2}$ were grown by the self-flux technique, as described in Ref. \onlinecite{Mal04}. X-ray diffraction and inductively coupled plasma atomic spectroscopy measurements showed no impurity phases and a chemical composition consistent with ideal stoichiometry.\cite{Cap05} The crystal structure of NaCu$_{2}$O$_{2}$  belongs to the \textit{Pnma} space group with an orthorhombic crystal structure and room-temperature lattice parameters $a=6.2087$\AA, $b=2.9343$\AA, $c=13.0648$\AA.\cite{Mal04} The unit cell is composed of two pairs of the edge-sharing $\rm Cu^{2+}O_2$ chains, running along the $b$ axis and shifted relative to each other by $b/2$, see Fig. 1. There are four Cu\(^{2+}\) ions per unit cell, and the density of Cu\(^{2+}\) ions is \(N_{\text{Cu}^{2+}}=1.68 \times 10^{22}\) cm\(^{-3}\). Within the single chain, the Cu\(^{2+}\)$-$Cu\(^{2+}\) distance is 2.934 \AA, and the Cu\(^{2+}\)$-$O$-$Cu\(^{2+}\) bond angle is 92.9\(^{\circ}\).

\begin{figure}[t]
\centering
\includegraphics[width=5.cm]{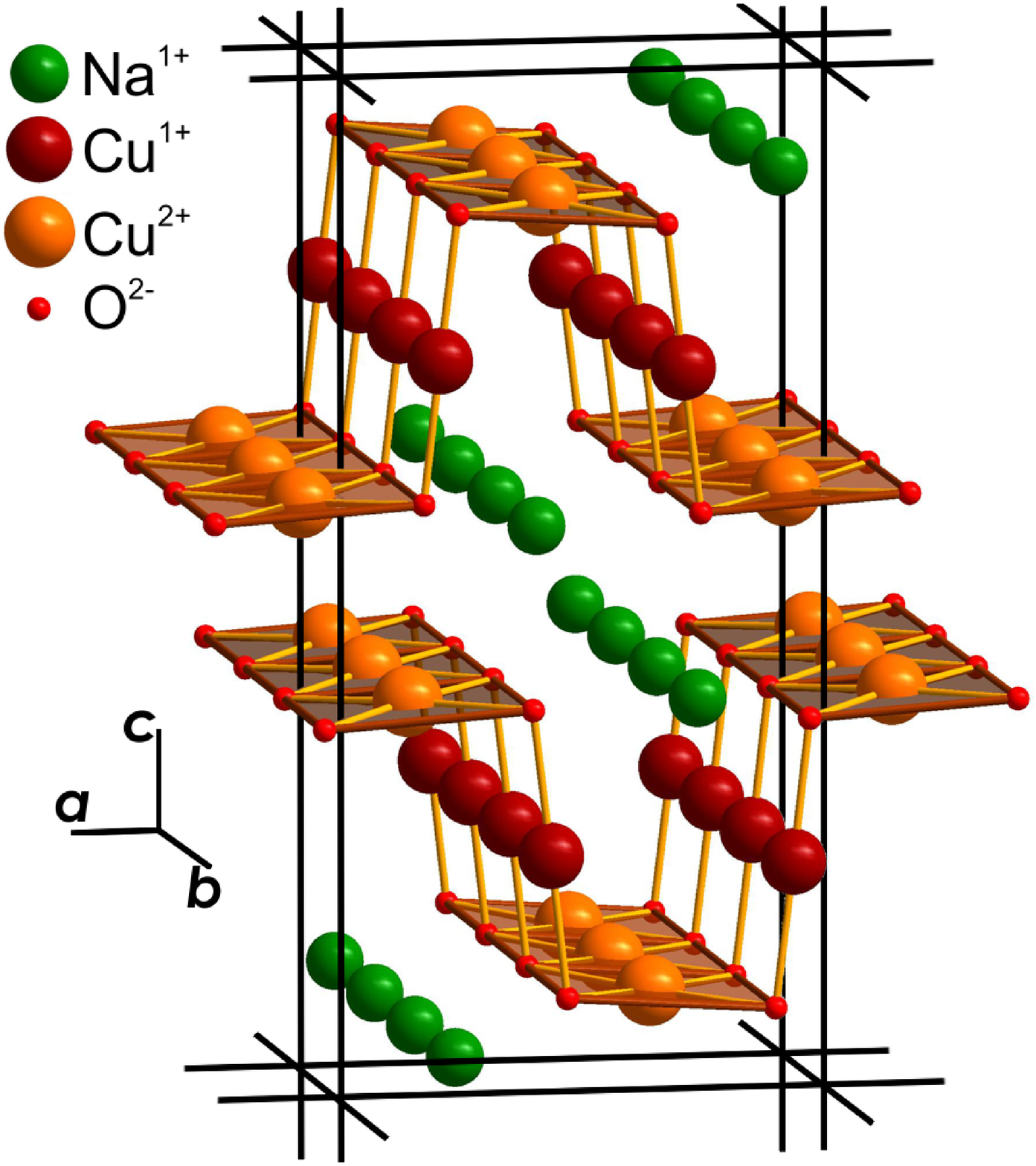}
\caption{Crystal structure of {\nco}.}
\label{fig0}
\end{figure}

Magnetic susceptibility data follow the Curie-Weiss law above $150$ K, with a Curie temperature of $-62$ K.\cite{Cap05} A broad maximum typical for low-dimensional magnets is observed  at $T = 52$ K, associated with short-range antiferromagnetic correlations within the chains. Low-temperature susceptibility and specific heat data indicate a magnetic phase transition at $T_{N}=11.5$ K.\cite{Lei10} Because of the 92.9\(^{\circ}\) Cu$-$O$-$Cu bond angle within the chains, the NN exchange integral is small ($J_{1}=-16.4$ K) and dominated by that between the next-nearest neighbors ($J_{2}=90$ K). Based on a refinement of single-crystal neutron diffraction data, an elliptical helix structure with alternating polarization planes
was proposed as the magnetic ground state.\cite{Cap10}

For our optical measurements a single crystal of NaCu$_{2}$O$_{2}$ with dimensions 5\(\times\)5\(\times\)0.2\ mm and the freshly cleaved \textit{ab} surface was used. The cleaving and mounting procedures were performed in argon atmosphere to prevent sample oxidation. The transfer of the cleaved and mounted sample into the cryostat was carried out in an argon-flooded glove bag. The sample was mounted on the cold finger of a helium-flow cryostat with a base pressure of \(2 \times 10^{-9}\) Torr at room temperature.
The ellipsometric measurements were performed with a rotating-analyzer type Woollam VASE variable angle ellipsometer.

Room temperature spectra of the ellipsometric angles $\Psi(\omega)$ and $\Delta(\omega)$ are shown in Fig. 2. The angles $\Psi$ and $\Delta$ are defined through the complex Fresnel reflection coefficients for light polarized parallel ($r_p$) and perpendicular ($r_s$) to the plane of incidence,
\begin{equation}
\tan \Psi e^{i \Delta}={r_p}/{r_s}.
\label{Eq:1}
\end{equation}
The data were measured at angles of incidence, $\varphi_i$, of 65\textordmasculine\ and 72.5\textordmasculine \ for sample orientations with the \textit{a} or \textit{b} axis in the plane of incidence, which denoted as \(\bot\) and \(\parallel \), respectively, corresponding to the orientation of the electric field with respect to the $\rm Cu^{2+}O_2$ chains. The observed dependence of the ellipsometric spectra on the sample orientation points to an intrinsic anisotropy of the optical response of {\nco}.

Ellipsometry yields the anisotropic frequency-dependent complex dielectric tensor,
\begin{equation}
\tilde \varepsilon(\omega) = \varepsilon_1(\omega)+i\varepsilon_2(\omega)=1+i\ 4\pi\tilde \sigma(\omega)/\omega,
\label{Eq:2}
\end{equation}
without the need for reference measurements or Kramers-Kronig transformations. A numerical regression procedure \cite{Elli} was applied to derive the principal components of the dielectric tensor from the ellipsometric data at different Euler angles. Since measurements of the optical response along the \textit{c} axis could not be obtained due to the thinness of the sample, spectra measured at two angles of incidence for each polarization were used in the fitting procedure.  The result of fits to the data are shown by thin lines in Fig. 2. The real and imaginary parts of the dielectric function for both polarizations resulting from the fit are drawn in Fig. 3. In this figure we also show the pseudo-dielectric function which is obtained by direct inversion of Eq. (1) for $\Psi(\omega)$ and $\Delta(\omega)$ measured at $\varphi_i= 65\textordmasculine$, assuming isotropic media with $\varepsilon_{xx}=\varepsilon_{yy}=\varepsilon_{zz}$. The effect of anisotropy mainly appears in the pronounced peaks in the spectral range 2.5 -- 4.0 eV. From this point on we will discuss only the corrected dielectric function.
\begin{figure}[t]
 \centering
 \includegraphics[width=8.0cm]{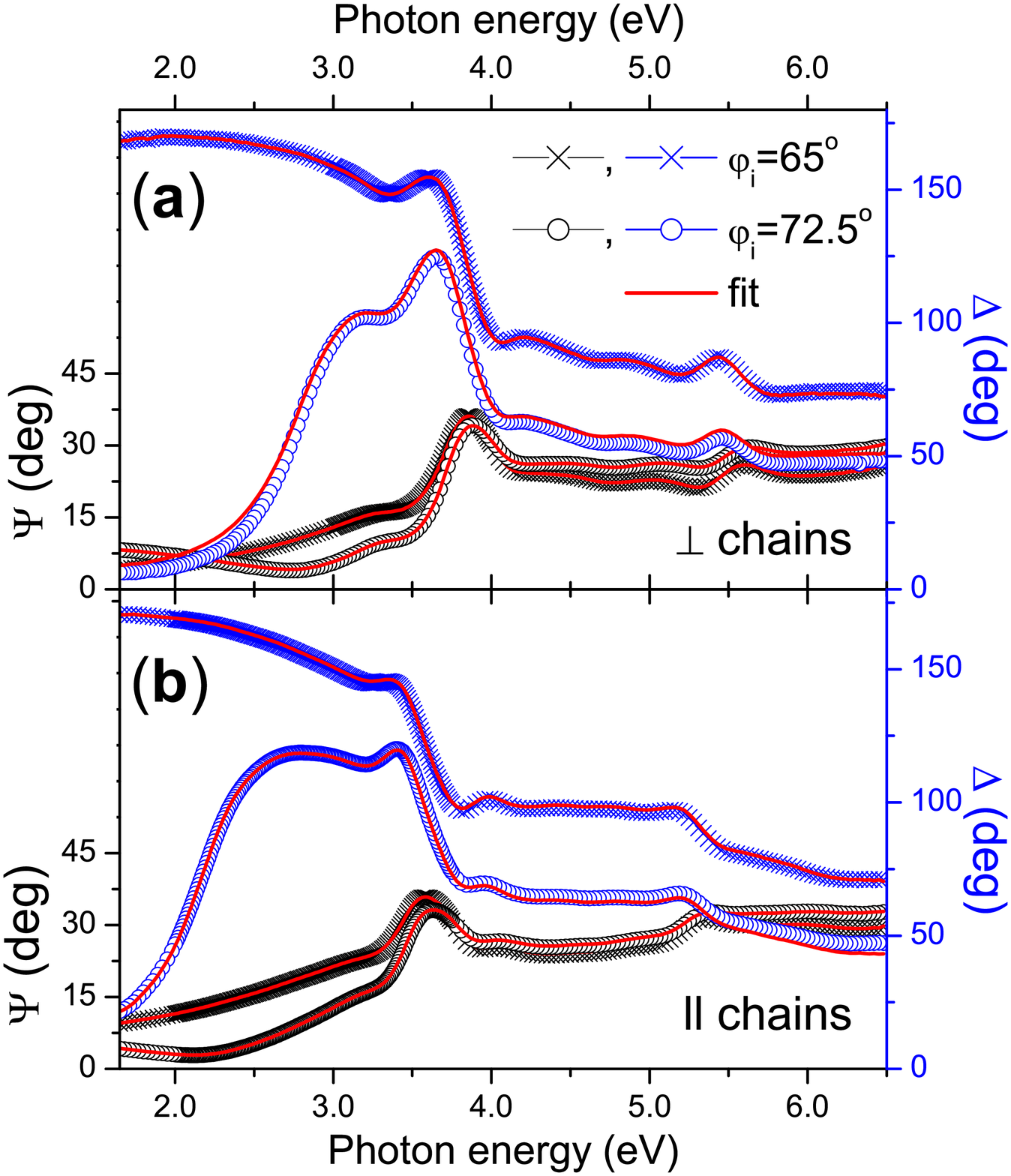}
 \caption{Ellipsometric angles $\Psi(\omega)$ and $\Delta(\omega)$ measured at room temperature for \textit{ab} sample surface orientations with the (a) \textit{a} or (b)  \textit{b} axis in the plane of incidence.}
 \label{fig2}
\end{figure}

\begin{figure}[t]
\centering
\includegraphics[width=8.0cm]{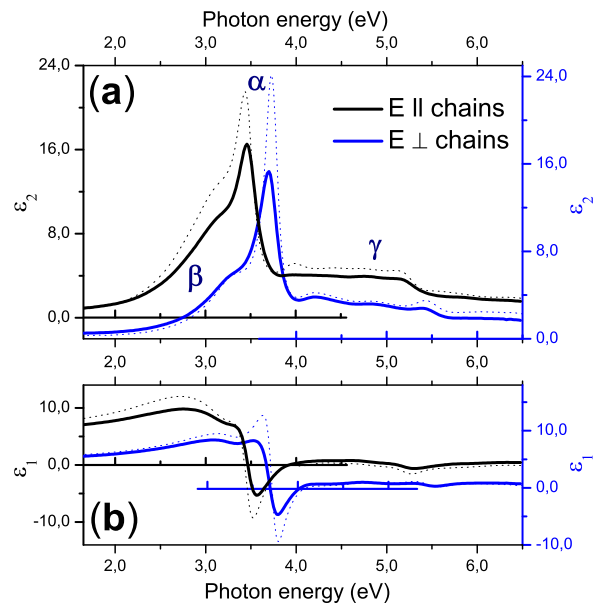}
\caption{Room temperature spectra of the real \(\varepsilon_{1}\) and imaginary  \(\varepsilon_{2}\) parts of the dielectric function for polarizations along and perpendicular to the chains of {\nco}. For polarization along the chains, the spectra are shifted by 2 for \(\varepsilon_{2}\) and 4 for \(\varepsilon_{1}\) for clarity. Dotted lines depict the pseudo-dielectric functions.}
\label{fig2}
\end{figure}

The optical spectra for both polarizations can be broken up into \(\alpha,\:\beta\) and \(\gamma\) zones, as shown in Fig. 3. The  \(\alpha\) zone comprises the spectral range of the strong and narrow absorption peaks located at 3.45 eV (along the chains) and 3.7 eV (perpendicular to the chains) that are dominating the spectra. The anomalous strength of these excitations  leads to negative values of \(\varepsilon_{1}\). The low-energy shoulders of these peaks near the insulating gap are marked as the \(\beta\) zone. The \(\gamma\) zone includes the broad and weak features above 4 eV that give rise to a rather flat and featureless optical response.

The temperature dependencies of the real parts of the optical conductivity $\sigma_{1}(\omega)$ and the real parts of the dielectric function  $\varepsilon_{1}(\omega)$ for both polarizations are plotted in Fig. 4. The strongest temperature effect on the optical spectra observed is a growth of the sharp 3.45 (3.7) eV peak for polarization along (perpendicular to) the chains with decreasing temperature. Against the background of this growth, some changes are apparent on the low-energy shoulders of the strong peaks. The low-temperature feature centered near 2.65 eV for polarization along the chains (see inset of Fig. 4(b)) is the most intriguing of these features.
\begin{figure}[t]
 \centering
 \includegraphics[width=8.0cm]{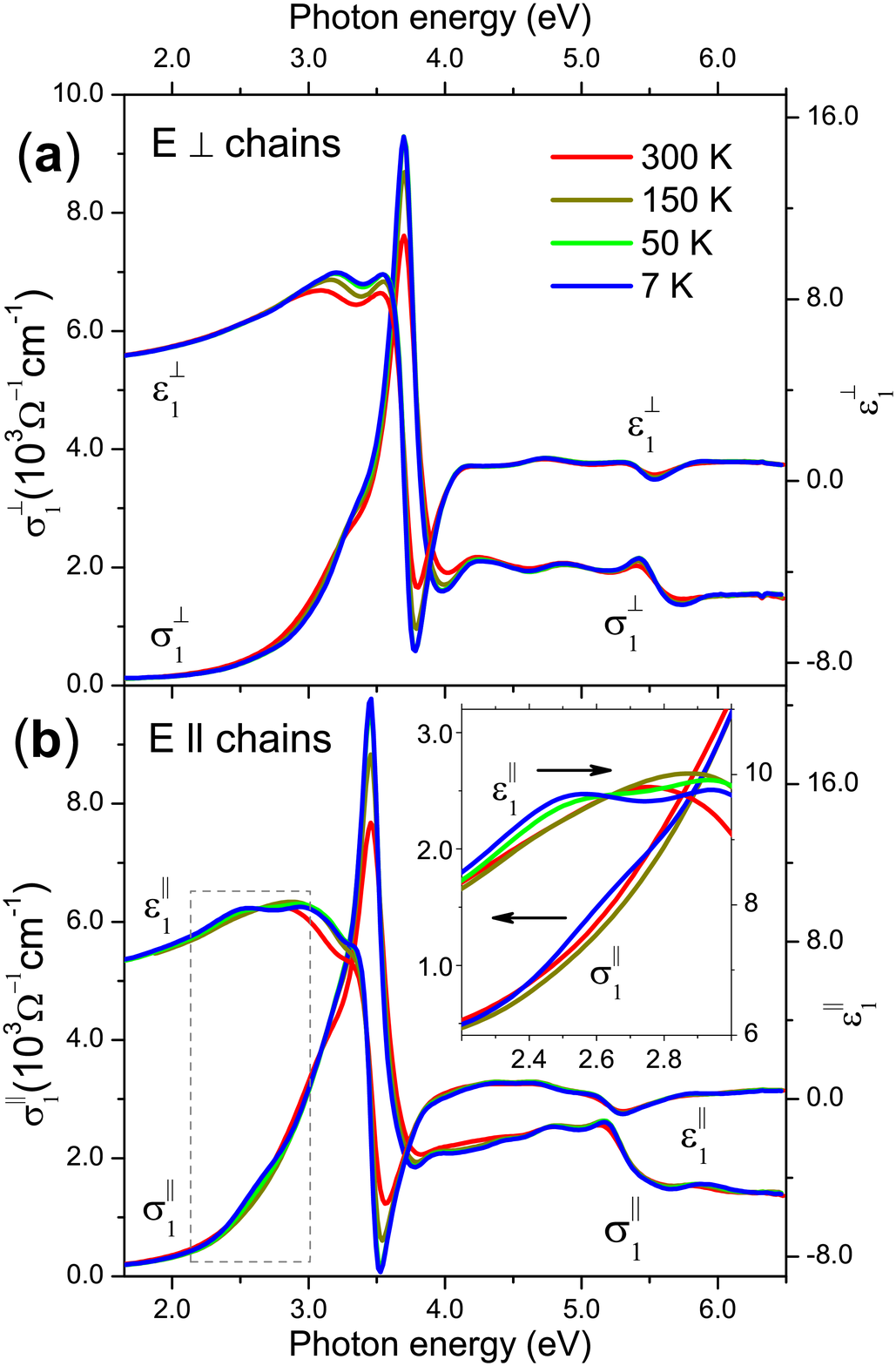}
 \caption{Real part \(\varepsilon_{1}\) of the dielectric function and real part \(\sigma_{1}\) of the optical conductivity for polarizations along and perpendicular to the chains of {\nco} measured at different temperatures. Inset: Magnified view of \(\sigma_{1}\) and \(\varepsilon_{1}\)  for polarization along the chains.}
\label{fig3}
\end{figure}

To study the temperature-driven changes in detail, temperature-difference spectra $\Delta \sigma_1(\omega,T)$ and $\Delta \varepsilon_1(\omega,T)$ with respect to 150 K  are plotted in Fig. 5. Perpendicular to the chains, ostensible changes in $\Delta \varepsilon^{\bot}_{1}(\omega)$ and $\Delta \sigma^{\bot}_{1}(\omega)$  are seen in the \(\alpha\) and \(\beta\) zones between 2.5 and 4.5 eV.  Alternating areas of positive and negative regions in the $\Delta \sigma^{\bot}_{1}(\omega) $ spectra within this spectral range are  mutually compensated. This observation indicates that the individual optical-band intensities within this spectral range are conserved, and that the spectral weight is not redistributed among these bands. Rather, the changes are  induced by concurrent narrowing of the pronounced band at 3.7 eV and the optical band located on its low-energy shoulder at 3.37 eV upon cooling.  Above 4.5 eV there are no other changes, except minor variations near 5.5 eV that can be attributed to a slight narrowing of the 5.4 eV band.

\begin{figure}[t!]
\centering
\includegraphics[width=8.0cm]{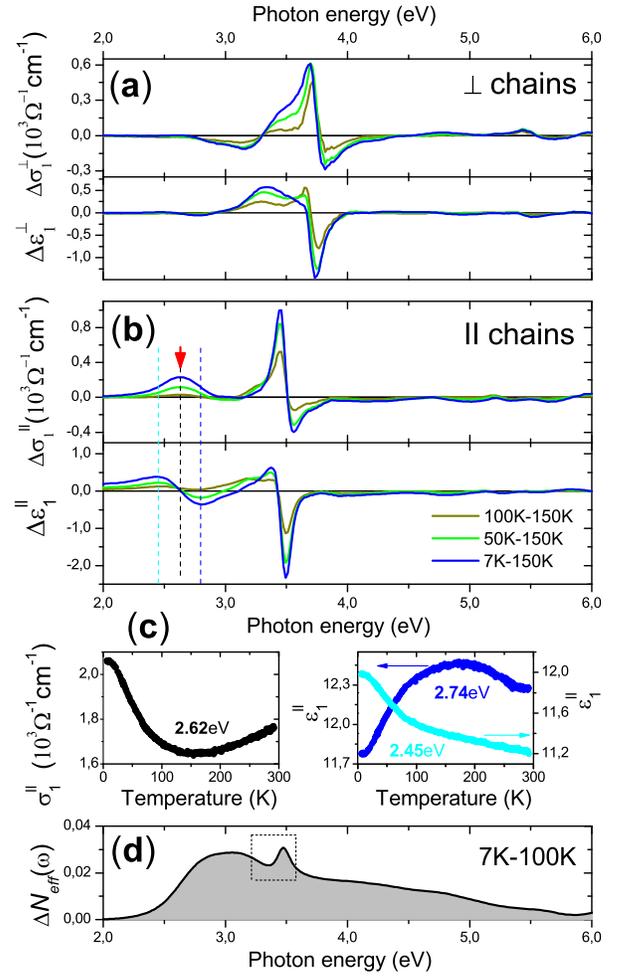}
\caption{Temperature-difference spectra $\Delta\sigma_1(\omega, T)=\sigma_1(\omega, T)-\sigma_1(\omega, 150$ K$)$
 and $\Delta\varepsilon_1(\omega, T)=\varepsilon_1(\omega, T)-\varepsilon_1(\omega, 150$ K$)$ of NaCu$_{2}$O$_{2}$ for polarizations (a) perpendicular to and (b) along the chains. (c) Temperature dependence of $\sigma^{\parallel}_1$ measured at $2.62$ eV and $\varepsilon^{\parallel}_1$ measured at $2.45$ and $2.74$ eV for polarization along the chains, as marked by vertical dashed lines in (b). (d) Spectral weight changes $\Delta N_{eff}(\omega)=N_{eff}(\omega, 7$ K$)-N_{eff}(\omega,100  $ K$)$ for polarization along the chains.}
\label{fig4}
\end{figure}

\begin{figure}[t!]
\centering
\includegraphics[width=8.0cm]{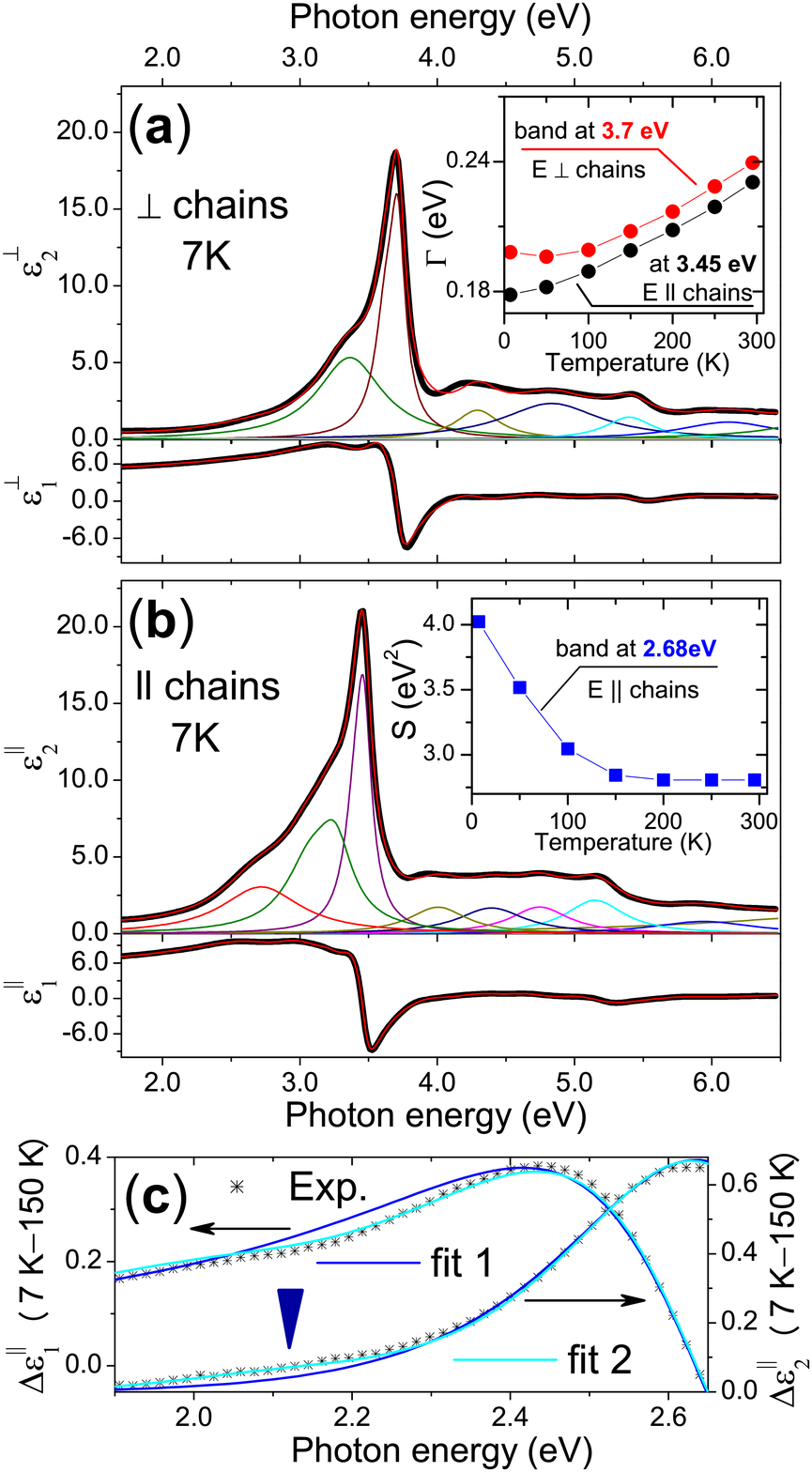}
\caption{Real \(\varepsilon_{1}\) and imaginary  \(\varepsilon_{2}\) parts of the dielectric function of {\nco} measured at 7K (black thick line) and dispersion analysis fit to the data (light thin line) for photon polarizations (a) perpendicular to and (b) along the $\rm Cu^{2+}O_2$ chains. Thin color lines represent the principal bands contributing to the optical response (see Table I), as derived from the dispersion analysis. (c) Temperature-difference spectra $\Delta\varepsilon_1(\omega)=\varepsilon_1(\omega, 7$ K$)-\varepsilon_1(\omega, 150$ K$)$ and $\Delta\varepsilon_2(\omega)=\varepsilon_2(\omega, 7$ K$)-\varepsilon_2(\omega, 150$ K$)$ for polarizations along the chains of {\nco}, and results of 'fit 1' and 'fit 2', as described in text. Insets: (a) Temperature dependence of the bandwidths of the Lorentz oscillators at 3.45 eV (black) and 3.7 eV (red) for polarizations along and perpendicular to the chains, respectively. (b) Temperature dependence of the oscillator strength of the 2.65 eV band for polarization along the chains.}
\label{fig5}
\end{figure}

Along the chains, the appearance of the feature marked with an arrow in Fig. 5(b) makes the temperature-difference spectra $\Delta \varepsilon^{\Vert}_{1}(\omega)$ and $\Delta \sigma^{\Vert}_{1}(\omega)$ qualitatively different from those for polarization perpendicular to the chains. The peak in $\Delta \sigma^{\Vert}_{1}(\omega)$ at 2.65 eV and anti-resonance with zero-crossing at the same energy in $\Delta \varepsilon^{\Vert}_{1}(\omega)$, which develop below 150 K, indicate an enhancement of the optical band upon cooling. The changes of the $\sigma^{\Vert}_{1}$ amplitude with temperature at the peak position for this band and $\varepsilon^{\Vert }_{1}$ right below and above, as marked with vertical dashed lines, were studied by temperature dynamic scans shown in Fig. 5(c). A Kramers-Kronig consistent enhancement of $\sigma^{\Vert}_{1}$ at 2.62 eV below 150 K, together with an upturn and a downturn in $\varepsilon^{\Vert }_{1}$ at 2.45 and 2.74 eV, respectively, confirm the pronounced intensity increase of the 2.65 eV band with decreasing temperature.

The transfer of spectral weight can be quantified by integrating the optical conductivity in terms of the effective charge density
\begin{equation}
\Delta N_{eff}(\omega,T)=\frac{2m}{\pi e^2 N_{\text{Cu}^{2+}}} \int_0^\omega \Delta\sigma^{\Vert}_1(\omega^{'},T)d\omega^{'},
\label{Eq:3}
\end{equation}
where $m$ is the free electron mass. Figure 5(d) shows the changes in the spectral weight below 100 K along the chains. The rise of the spectral weight at low energies is due to the band at 2.65 eV. Above 3 eV this growth is compensated by a loss in the spectral weight of the higher-energy bands, following the optical sum rule. The feature marked with a rectangle in Fig. 5(d) originates from the narrowing of the strong optical band at 3.45 eV with decreasing temperature and indicates that the spectral weight is retained within this band.

To separate the contributions from different optical bands to the dielectric function spectra for both polarizations,  a classical dispersion analysis was performed. A minimum set of Lorentzian oscillators, with one high-energy oscillator beyond the investigated spectral range, was introduced to represent a dielectric function in the form
\begin{equation}
\varepsilon(\omega) =\varepsilon_{\infty}+ \sum_j\frac{S_j}{\omega_j^2-\omega^2-\text{i}\omega\Gamma_j},
\label{Eq:4}
\end{equation}
where $\omega_j$, $\Gamma_j$, and $S_j$ are the peak energy, width, and
oscillator strength of the $j$th oscillator, and $\varepsilon_{\infty}$ is the high energy core contribution to the dielectric function. The parameters  of the individual complex Lorentzian oscillators were derived with high accuracy and reliability by simultaneously fitting to $\varepsilon_{1}(\omega)$ and $\varepsilon_{2}(\omega). $ They are listed in Table I.

Figures 6(a) and 6(b) represent the principal bands contributing to the dielectric response at 7 K for polarizations along and perpendicular to the chains.
\begin{table}[b!]
\caption{\label{tab:table1}Parameters of Lorentz oscillators resulting from a dispersion analysis in polarizations perpendicular and (along) the chains in NaCu\(_{2}\)O\(_{2}\) measured at \(T=7\) K. \(\varepsilon_{\infty}\)=2.00, (\(\varepsilon_{\infty}\)=1.77).  }
\begin{ruledtabular}
\begin{tabular}{lrrr}
& $\omega_j$(eV)&$S_{j}$(eV\(^{2}\))&$\Gamma_j$(eV)\\
\hline
        &   (2.65) &  (4.02) &  (0.60)\\
\(\beta\)&  3.37 (3.08) & 10.3 (7.40) & 0.57 (0.52)\\
        & (3.26) &  (4.99) &  (0.33)\\
\hline
\(\alpha\)&3.63 (3.40) & 5.53 (3.70) & 0.19 (0.19)\\
         &3.71 (3.46) & 5.23 (5.98) & 0.12 (0.14)\\
\hline
         &4.30 (4.01) & 3.14 (3.52) & 0.39 (0.51)\\
         &4.85 (4.41) & 9.88 (4.47) & 0.87 (0.58)\\
\(\gamma\)          &(4.75) &  (4.29) &  (0.52)\\
         &5.40 (5.16) & 3.20 (5.86) & 0.43 (0.52)\\
         & 6.14 (5.96) & 6.03 (4.26) & 0.87 (0.89)\\
         &7.07 (7.73) & 13.0 (35.3) & 0.88 (3.02)\\
\end{tabular}
\end{ruledtabular}
\end{table}
According to the dispersion analysis, the $\alpha$ zone of the spectra is composed of two  narrow bands at 3.63 (3.40) and 3.71 (3.46) eV for polarization perpendicular to (along) the chains, giving in sum the extremely strong asymmetric absorption band at 3.7 (3.45) eV. The \(\gamma\) zone of the spectra is composed of a series of excitation bands. Along the chains, four nearly equally spaced optical bands (at 4.01, 4.41, 4.75 and 5.16 eV) can be recognized. Perpendicular to the chains, the dispersion analysis gives a broad band at 4.85 eV. However, this band can be split into two subbands that are surrounded by excitations at 4.30 and 5.40 eV, forming a series of excitations similar to that for polarization along the chains. The \(\gamma\) regions in both polarizations end in a broad weak band near 6 eV and one optical band beyond the investigated spectral range. Only one band located at 3.37 eV is lying in the \(\beta\) zone for polarization perpendicular to the chains. Along the chains this band is resolved into two sub-bands at 3.08 and 3.26 eV. In addition, there is an extra band at 2.65 eV, which is the lowest-energy excitation along the chains.

The temperature evolution of the bands at 3.7 eV perpendicular to the chains and at 2.65 and 3.45 eV along the chains can be also obtained from the dispersion analysis. The total oscillator strength of the 3.7 (3.45) eV $\alpha$ bands is 10.76 (9.68) eV\(^{2}\). While the spectral weights of these bands are conserved and independent of temperature, their widths decrease gradually with decreasing temperature, as shown in the inset of Fig. 6(a). The dispersion analysis demonstrates a prominent strengthening of the optical band at 2.65 eV along the chains upon cooling below 150 K (see the inset of Fig. 6(b)), which is in full agreement with the measured dynamical scans shown in Fig 5(c).

Close inspection of the temperature-difference spectra $\Delta \varepsilon^{\parallel}_{1}(\omega)$  and $\Delta \varepsilon^{\parallel}_{2}(\omega)$ in the spectral region below 2.5 eV highlighted in Fig. 6(c) reveals a deviation of the fit incorporating the bands listed in Table I (labeled as 'fit 1') from the experimental data indicated by stars. The deviation in both $\Delta \varepsilon^{\parallel}_{1}(\omega)$  and $\Delta \varepsilon^{\parallel}_{2}(\omega)$ spectra can be removed by introducing an additional band at 2.1 eV into the fit (labeled as 'fit 2') in Fig. 6(c). However, the weakness of the additional absorption band at 2.1 eV did not allow us to study its temperature dependence.

\begin{figure}
\includegraphics[width=8.0cm]{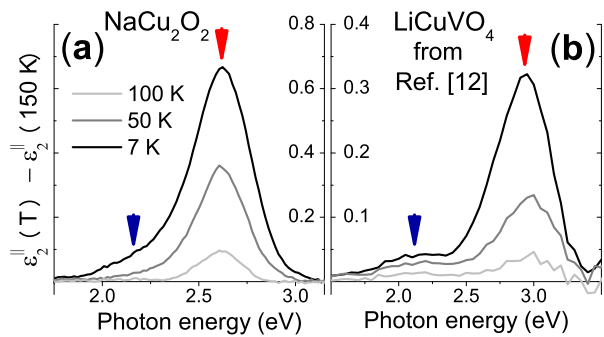}
\caption{Temperature-difference spectra $\Delta\varepsilon_2(\omega)=\varepsilon_2(\omega, T)-\varepsilon_2(\omega, 150$ K$)$ along the chains in (a) {\nco} and (b) {\lcvo} at low energies. The red and blue arrows mark the positions of the two excitonic modes.}
\label{fig6}
\end{figure}

\section{\label{sec:Discussion}discussion}

\subsection{Optical response of the $\rm Cu^{2+}O_2$ chains}

The optical response for both polarizations is composed of similar sets of oscillators, as revealed by the dispersion analysis. The only qualitative difference is the temperature-dependent band at 2.65 eV, which is clearly resolved at low temperatures along the chains. Comparison of its temperature evolution (Fig. 5(c) and inset of Fig. 5(b)) with magnetic susceptibility data \cite{Lei10} points to the spin-controlled behavior of this band.  When spin-correlations appear within the  $\rm Cu^{2+}O_2$ chains below 150 K, the magnetic susceptibility data deviate from Curie-Weiss behavior.\cite{Mal04} Concurrently with the appearance of spin correlations, the intensity of the 2.65 eV band is enhanced, while it is virtually temperature independent above 150 K.

The dispersion analysis at $T=7$ K picked out a low-energy side band at 2.1 eV, see Fig. 6(c), apparent at 7 K. Therefore, the lowest-energy excitation along the $\rm Cu^{2+}O_2$ chains has a doublet structure, as shown in Fig. 7(a). Its characteristics including the energy positions, the partial spectral weights of the  individual bands, and their evolution with temperature, are similar to those of the two-peak structure centered at 2.15 and 2.95 eV observed along the chains in {\lcvo}.\cite{Mat09} This analogy indicates that these excitations have the same origin in both materials. Following arguments presented in Ref. \onlinecite{Mat09}, we therefore ascribe the double-peak structure observed in {\nco} to an exciton doublet formed by $d^9 L_h$ and $d^{10}$ states generated by the NN ($t_1$) and NNN ($t_2$) hopping of electrons along the chains that emerges at $U-V/l$ as a consequence of the long-range Coulomb interaction between electrons on neighboring sites $l=1,2$.\cite{Mat09,Hor05} The parameters characterizing the local Hubbard interaction \(U=3.2\) eV and the long range Coulomb repulsion \(V=1.1\) eV can be obtained from the energies of these bands. Within this model the spectral weight for the first and second excitons are directly related to the corresponding spin correlations: \cite{Mat09}
\begin{equation}
N_{eff}^{(l)}=-\frac{2m}{\hbar^2} d^2_l J_l
\langle \vec{S_i}\cdot \vec{S}_{i+l}-1/4 \rangle,\ l=1,2,
\label{Eq:5}
\end{equation}
where $d_l$ is the hopping length ($d_2=2d_1$) and $J_l\simeq 4 t_l^2/(U-V_l)$.
The large value of the antiferromagnetic coupling constant,
$J_2$, leads to larger variation in $N_{eff}^{(2)}$ as a function of temperature compared to $N_{eff}^{(1)}$. The oscillator strengths of the individual bands in the difference spectra of $\Delta \varepsilon^{\parallel}_{2}(\omega, 7$ K$)-\Delta \varepsilon^{\parallel}_{2}(\omega, 150$ K$)$ in Fig. 7 implies a ratio $N_{eff}^{(2)}/N_{eff}^{(1)} \approx$ 24 in {\nco} and 11.5 in {\lcvo}. On the other hand, a ratio between the NN and NNN exchange integrals extracted form neutron scattering data is  $\alpha\sim$ 5.5 in {\nco}\cite{Cap05}, compared with the reduced ratio $\alpha \sim$ 2.7 in {\lcvo}.\cite{End05} A precise quantitative determination of the exchange couplings in {\lcvo} is disputed.\cite{End10,Dre11,End11,Koo11} Nevertheless, the qualitative agreement between $N_{eff}^{(2)}/N_{eff}^{(1)}$ and \(\alpha\) ratios for two compounds validates Eq. (5) for the spectral weight of the exciton bands.

Specifically, the spin correlation functions in Eq.(5) are determined by the Heisenberg Hamiltonian of the underlying frustrated $J_1$-$J_2$ spin chain:\cite{Mat09}
\begin{equation}
H_s=\sum_l J_l \sum_i(\vec{S_i}\cdot \vec{S}_{i+l}-1/4)+
J_1^F \sum_i \vec{S_i}\cdot \vec{S}_{i+l},
\label{Eq:6}
\end{equation}
with the total NN exchange integral determined by the balance of two opposing contributions, $J_1^{tot}=J_1+J_1^F$. Apart from the antiferromagnetic superexchange integrals $J_1$ there is a substantial ferromagnetic coupling $J_1^F$ that originates from a two hole excitation on the oxygen site that is not included in Eq.(5). This interaction, however, plays an important role for the frustrated magnetism of these spin chains. We may use now the additional experimental information of the relative weights $N^{(2)}_{eff}/N^{(1)}_{eff}\approx 24$ at low temperature together with the total exchange integrals obtained for {\nco} \cite{Cap05} $J_1^{tot}=-16 $ K and $J_2^{tot}=J_2=90$ K to determine $J_1$ and $J_1^F$ separately. With help of an estimate of the correlation functions $\langle\vec{S_i}\cdot \vec{S}_{i+l}\rangle \simeq 0\ (-0.4)$ for $l=1\ (2)$ at zero temperature we find $J_1\simeq 39$K, which implies for the ferromagnetic contribution $J_1^F\simeq -55$K.

\subsection{In-plane anisotropy and anomalous absorption}

The in-plane dielectric function of {\nco} measured with photon polarization parallel to the $\rm Cu^{1+}$ planes exhibits strong absorption bands at 3.45 eV and at 3.7 eV perpendicular and parallel to the chains, respectively, which are unusual for strongly correlated electron systems. The energy difference of 0.25 eV between their peak positions is a basic tendency also for other bands marked out by the dispersion analysis. The unique exception is the 2.65 eV band, which is apparent only along the chains, as discussed in detail above. Since compounds with mixed-valent Cu atoms are scarce, these experimental data are of special interest for studying the low-energy electronic excitations in cuprates.

One of these mixed-valent compounds is the parent compound of a family of high-$T_c$ superconductors {\ybco}, which shares some of its structural units with {\nco}: Cu\(^{2+}\)  ions are centered in the CuO\(_{4}\) plaquettes forming the CuO\(_{2}\) conducting planes, and Cu\(^{1+}\) ions construct the O-Cu\(^{1+}\)-O dumbbells. Ellipsometric measurements with light polarized within the \textit{ab} plane along the \textit{a}-axis showed that the dielectric function of {\ybco} exhibits a sharp and intense peak at 4.1 eV, \cite{Kir89} which disappears with increasing oxygen content.\cite{Kel89} Local density approximation (LDA) calculations assign the dominant absorption peak to intra-ionic transitions within Cu\(^{1+}\) ions of the O-Cu\(^{1+}\)-O complexes. \cite{Kir89} The initial states were assigned to Cu\(^{1+}\) \(3d_{3z^{2}-1}\) orbitals, and the final states consist of bonding combinations of Cu\(^{1+}\) \(4p_{x}\) orbitals and Ba 5$d$ and 4$f$ orbitals. Although the Cu\(^{1+}\) \(3d_{3z^{2}-1}\) \(\rightarrow\) Ba transitions contribute only slightly to the optical matrix elements. Nonetheless, the analogy to the O-Cu\(^{1+}\)-O transitions in \nco\ is not complete, because in this compound the oxygen atoms are shared with the Cu$^{2+}$-O$_2$ chains.

To our knowledge, the highest value of \(\varepsilon_{2}(\omega)\) among all transition-metal oxides has been observed in {\lco}, a compound that is isostructural and isoelectronic to {\nco}. In a recent ellipsometry study, Pisarev {\it et al.} reported that the real part of the dielectric permittivity of {\lco} exhibits an extremely strong and narrow absorption peak at 3.27 eV. Overall, the in-plane dielectric response resembles our measured data and is composed of a similar set of Lorentz oscillators. However, Li-Cu chemical substitution, a 10\% nonmagnetic LiCuO impurity phase and the twinned nature of {\lco} crystals \cite{Mas04, Zvy02} obliterate the dielectric anisotropy in the \textit{ab} plane. Reviewing the optical data of the large family of cuprates with O-Cu\(^{1+}\)-O complexes, Pisarev {\it et al.} \cite{Pis06} pointed out a relationship between the Cu\(^{1+}\)-Cu\(^{1+}\) and Cu\(^{1+}\)-O\(^{2-}\) bond lengths and the position and intensity of the sharp peak, which implies that this feature originates from the dumbbells. Arguing on the basis of the peculiar intensity of the 3.27 eV band, the authors proposed an exciton model and attributed the observed anomaly to a strong Cu\(^{1+}\) 4$p$-3$d$ electron-hole interaction along with a strong crystal-field splitting of the excited states.

In order to interpret the observed anisotropy and clarify the origin of the intensive absorption peak, we performed band structure calculations of {\nco} along the high symmetry directions of the Brillouin zone.

\begin{figure}
\includegraphics[width=8.5cm]{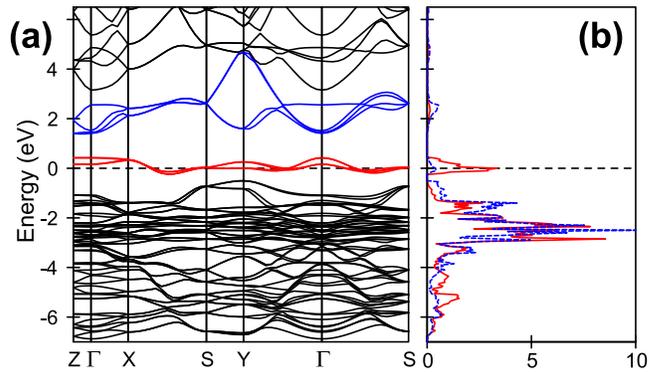}
\caption{(a) Energy band dispersion calculated by the spin-restricted LSDA approximation for {\nco} with partially filled Cu\(^{2+}\) 3\textit{d} (red) and unoccupied Cu\(^{1+}\) hybrid 4\textit{p} and 3\textit{d}\(_{3z^{2}-1}\) (blue) states. (b) DOS projected on the Cu\(^{2+}\) 3\textit{d}  (red) and Cu\(^{1+}\) 3\textit{d}  (blue) atomic states of {\nco}. The Fermi level is at zero energy.}
\label{fig7}
\end{figure}
\begin{figure}
\includegraphics[width=7.0cm]{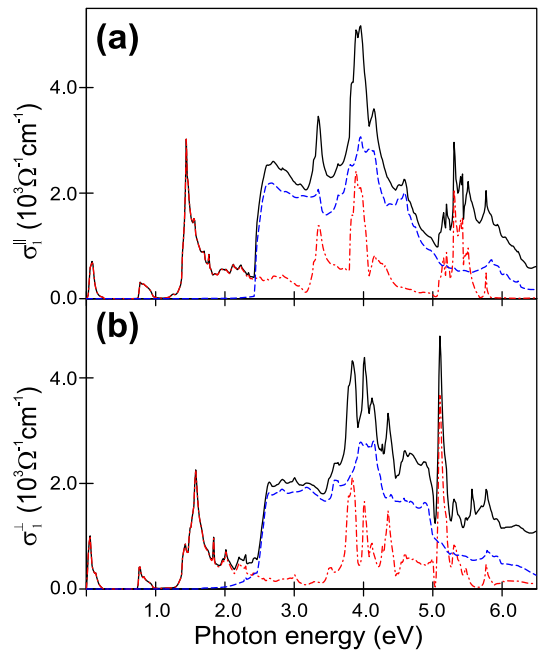}
\caption{Calculated diagonal elements of the (a) \sryy$(\omega)$ and (b) \srxx$(\omega)$ optical conductivity of {\nco} (black solid line) and its decomposition into transitions to the Cu\(^{2+}\) (red dash-dotted line) and the Cu\(^{1+}\) (blue dashed line) final states. }
\label{fig8}
\end{figure}
\begin{figure*}[ht]
\includegraphics[width=14.5cm]{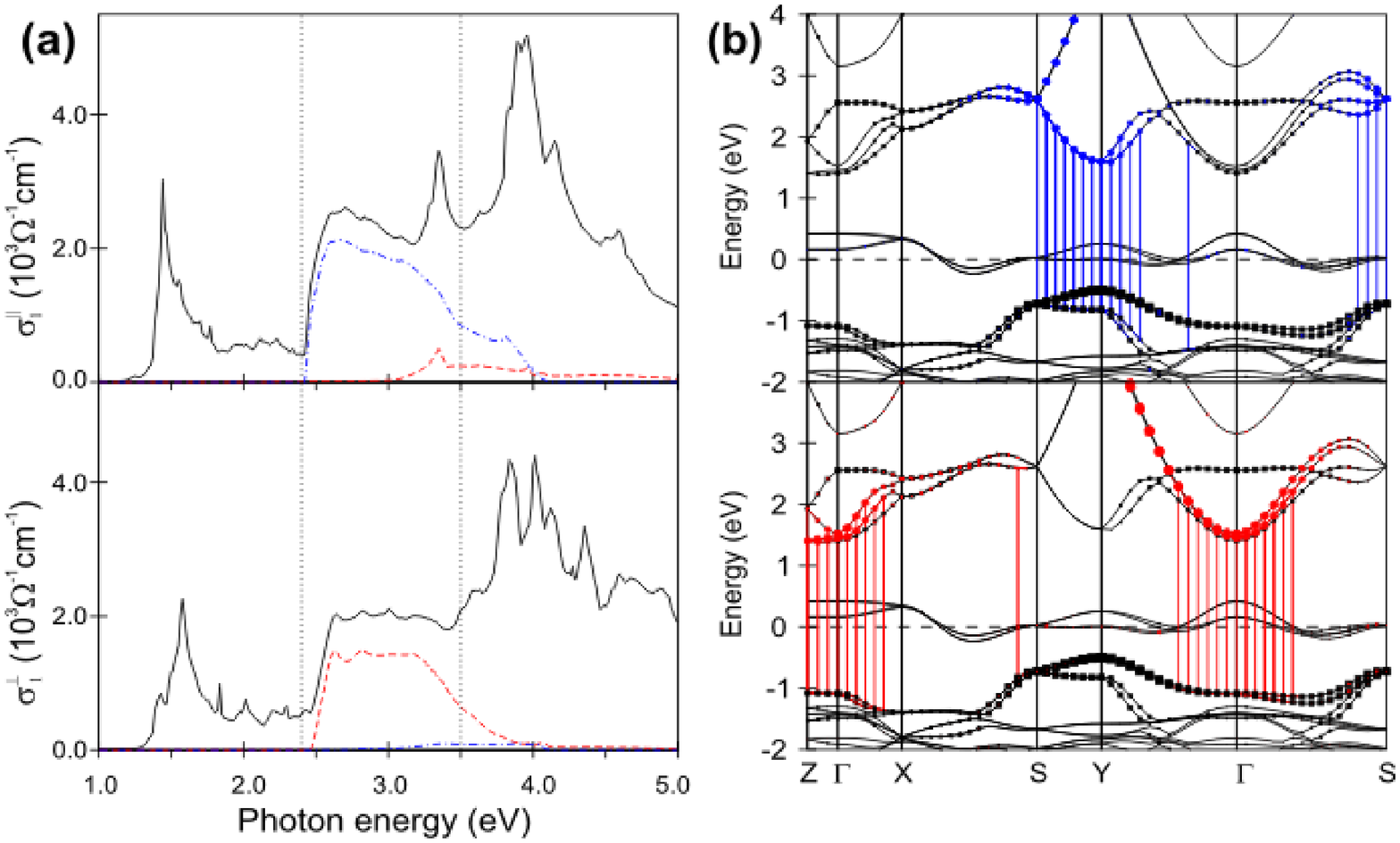}
\caption{(a) Partial contribution of transitions from the occupied Cu$^{1+}$ 3$d_{3z^{2}-1}$ to the unoccupied Cu$^{1+}$ 4$p_{x}$ (blue dash-dotted line) and  4$p_{y}$ (red dashed line) states to \sryy$(\omega)$ and \srxx$(\omega)$ within spectral range 2.35$-$3.5 eV. (b) The belts of transitions from Cu$^{1+}$ 3$d_{z^{2}-1}$ to 4$p_{x}$ (blue) and to 4$p_{y}$ (red) states, partially contributing \sryy$(\omega)$ and \srxx$(\omega)$ in (a) and the energy band dispersion in fat-band representation. The size of black squares, blue and red circles in (b) is proportional to the partial weights of the Cu$^{1+}$ 3$d_{z^{2}-1}$, 4$p_{x}$ and 4$p_{y}$ states in the Bloch wave function, respectively.}
\label{fig9}
\end{figure*}
\begin{figure*}[ht]
\includegraphics[width=14.5cm]{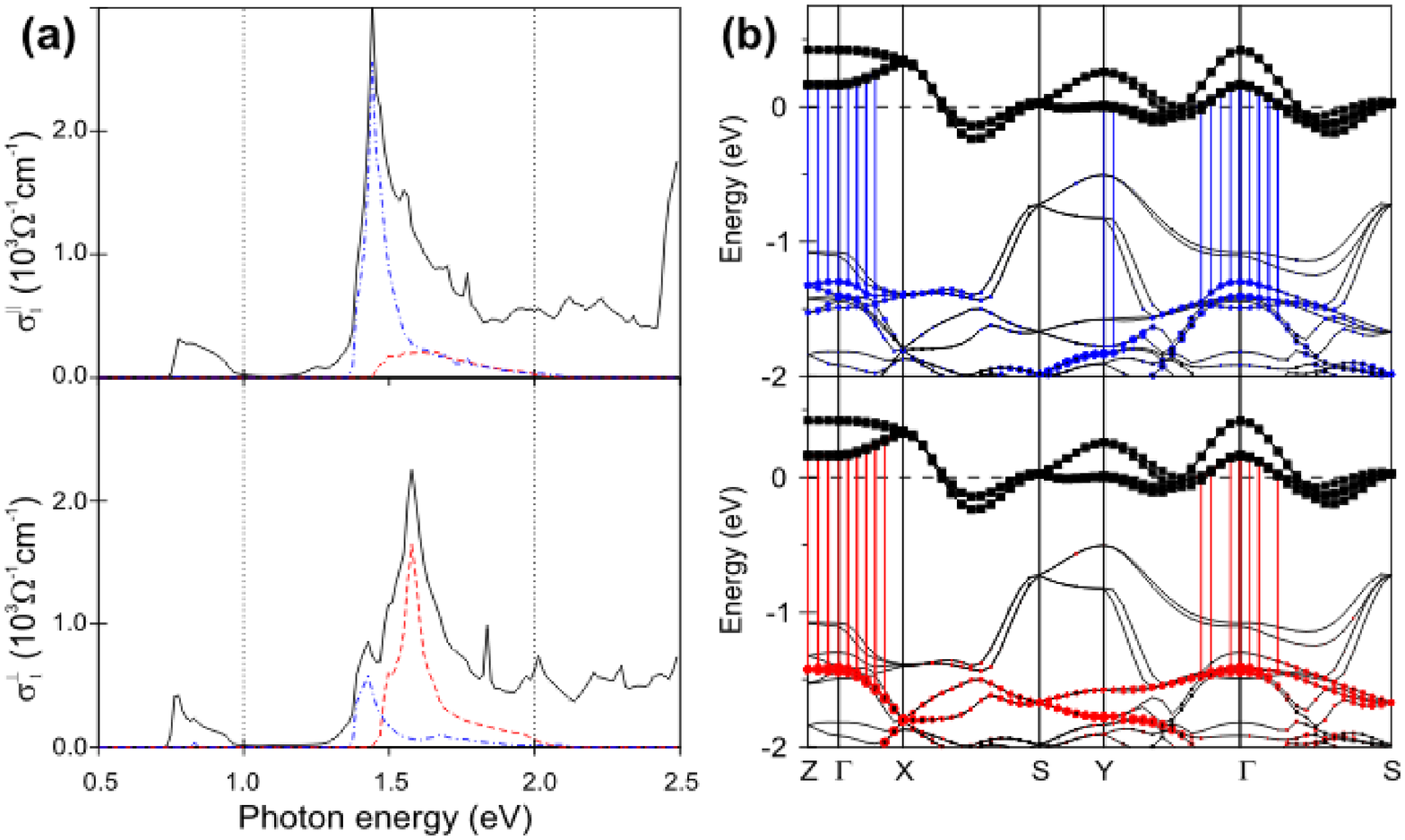}
\caption{(a) Partial contribution of transitions from the occupied Cu\(^{1+}  \,\text{3}d_{xz}\) (blue dash-dotted line) and and 3\(d_{yz}\) (red dashed line) to the unoccupied Cu\(^{2+}\) 3\(d_{xy}\) states to \sryy$(\omega)$ and \srxx$(\omega)$ within the spectral range from 1$-$2 eV. (b) The belts of transitions from Cu\(^{1+}  \,\text{3}d_{xz}\) (blue) and Cu\(^{1+}  \,\text{3}d_{yz}\) (red) to Cu\(^{2+}\) 3\(d_{xy}\) states, partially contributing to
\sryy$(\omega)$ and \srxx$(\omega)$ in (a) and the energy band dispersion in fat-band representation. The size of black squares, blue and red circles is proportional to the partial weights of the Cu\(^{2+}  \,\text{3}d_{xy}\), Cu\(^{1+}  \,\text{3}d_{xz}\) and  Cu\(^{1+}  \,\text{3}d_{yz}\) states in the Bloch wave function, respectively.}
\label{fig10}
\end{figure*}
\begin{figure}[hb]
\includegraphics[width=7.0cm]{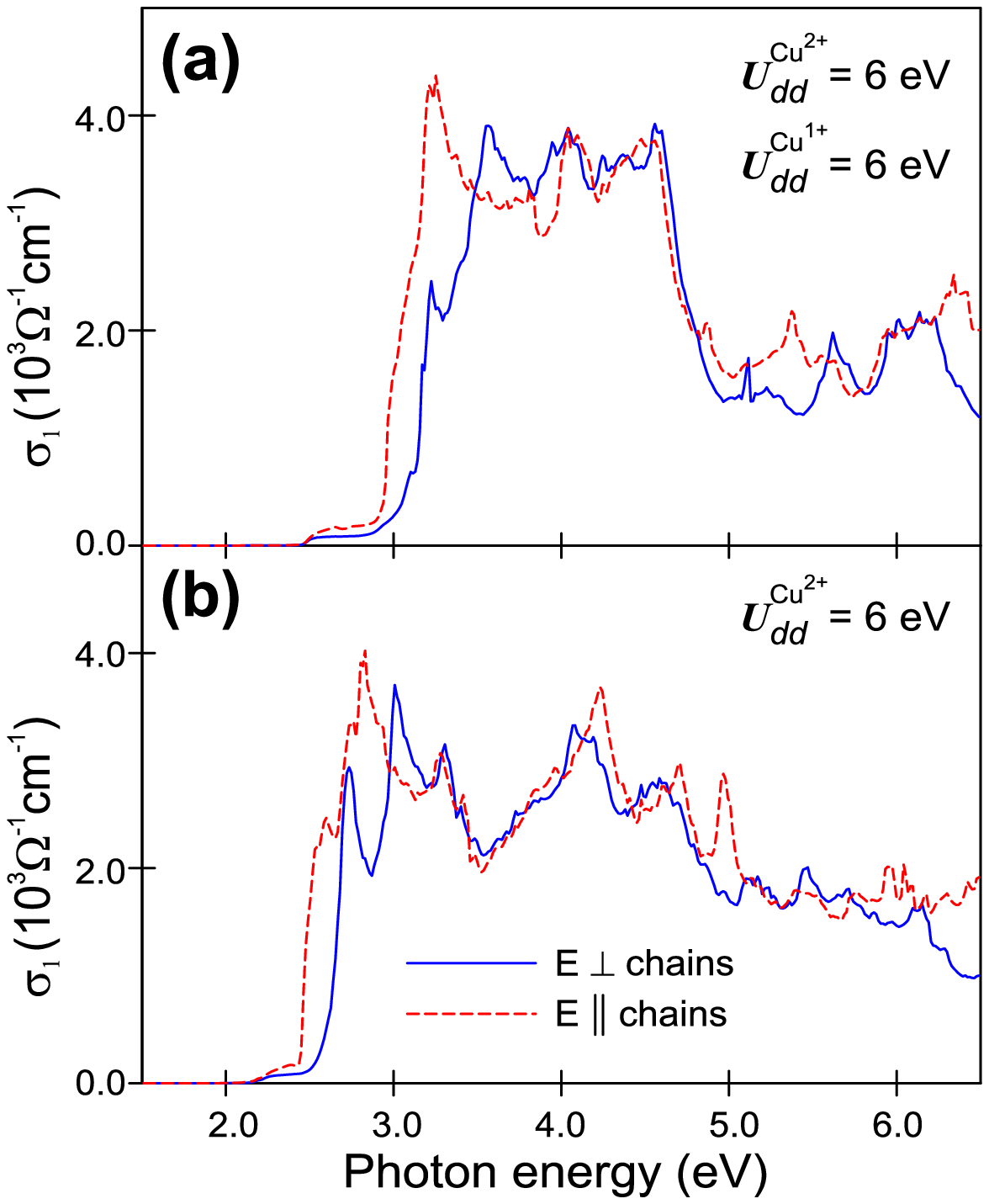}
\caption{Optical conductivities \sryy$(\omega)$ and (b) \srxx$(\omega)$ calculated with $U_{dd}$=6 eV acting (a) on the Cu$^{2+}$ and Cu$^{1+}$ $d$  states  and (b) on the Cu$^{2+}$ $d$ states only.}
\label{fig11}
\end{figure}

\subsection{Band Structure Calculations}

Band structure calculations were performed using a linear-muffin-tin orbital method in the atomic sphere approximation \cite{And75} within the local spin density approximation (LSDA) starting from the known crystal structure of {\nco}.

The results of the spin-restricted LSDA calculations along with the resulting partial electronic densities of states (DOSs) are shown in Fig. 8. The Cu\(^{1+}\) and  Cu\(^{2+}\) 3\textit{d} DOS are shown in Fig. 8(b), as they are the main contributors to the total DOS. Since the LSDA results in a metallic state, narrow partially filled bands corresponding to Cu\(^{2+}\) 3\textit{d} orbitals with \textit{xy, xz} symmetry cross the Fermi level and give a DOS peak at \(E_{F}\). The valence band between $-$4 eV and $-$1 eV below the Fermi energy \(E_{F}\) is a mixture of Cu\(^{1+}\) and Cu\(^{2+}\) 3\textit{d} states of different symmetries strongly hybridized with O 2\textit{p} states. The modestly intense band at near 2 eV above the Fermi energy shown by blue lines in Fig. 8(a) is mainly formed by unoccupied 4\textit{p} and 3\textit{d}\(_{3z^{2}-1}\) states of Cu\(^{1+}\).

The calculated optical conductivity spectra \sryy$(\omega)$ and \srxx$(\omega)$ obtained from the band structure are shown in Fig. 9. The sums of all the interband transition contributions to the optical conductivity (the total optical conductivity) are shown by black solid lines. Blue dashed and red dash-dotted lines indicate the spectral weights associated with transitions to the unoccupied Cu\(^{1+}\) and Cu\(^{2+}\) states, respectively. The calculated optical response below 2.4 eV originates entirely from transitions involving Cu\(^{2+}\) final states, as discussed in detail below.

The calculated spectra above 2.4 eV, contributed mostly by transitions into the Cu\(^{1+}\) states, conform to the experimentally observed response in the \(\gamma\) region, which exhibits two dominant bands at 3.45 and 3.7 eV ($\alpha$ region Fig. 3) on top a broad background ($\beta$ and $\gamma$ regions). While the dispersion analysis of the experimental data separates the \(\gamma\) zone into bands with half-widths $\sim $ 0.5 eV (Fig. 6(a) and 6(b)), the LSDA gives sharper structures. However, within this spectral range, the average calculated optical conductivity $\sim$ 2.5$\times10^{3}\ \Omega^{-1}\text{cm}^{-1}$ is in agreement with the measured data. The intensities of both calculated and experimental spectra are reduced above 5 eV.

To find out the origin of the observed in-plane anisotropy, appearing as a shift of the spectra by 0.25 eV, we analyzed the interband transitions contributing to the optical conductivity above 2.4 eV. The decomposition of the calculated \srxx$(\omega)$ and \sryy$(\omega)$ spectra into separate transitions revealed that the Cu\(^{1+}\) 3\(d_{3z^{2}-1}\) orbitals serve as \textit{initial} states for transitions within the spectral range 2.4$-$3.5 eV in Fig. 10(a). The \textit{final} states involved are of  Cu\(^{1+}\) 4\(p\) character with \textit{x} and \textit{y} symmetries. The respective transitions and dispersions of bands in a fat-band representation are illustrated in Fig. 10(b). As shown in Fig. 10(a),  the \sryy$(\omega)$ spectrum around 3 eV  is, mainly, formed by transitions into the Cu\(^{1+}\) 4\(p_{x}\) states (blue dash-dotted lines), while transitions into the 4\(p_{y}\) states (red dashed lines) dominantly form the  \srxx$(\omega)$ spectra. Because of the large dispersion of these states, the transitions give a rather flat shape of the optical conductivity in this spectral range and spread out up to 5 eV. Above 3.2 eV the calculated spectra are formed by the Cu\(^{1+}\) 3\textit{d} unoccupied states and O 2\textit{p} \(\rightarrow \) Cu\(^{2+}\) 3\textit{d} interband transitions, contributing equally to both \srxx$(\omega)$ and \sryy$(\omega)$.

Let us turn now to the interband transitions associated with the Cu\(^{2+}\) 3\textit{d} \textit{final} states, giving, in particular, the sharp peaks near 1.5 eV in Fig. 9. The decomposition into partial contributions reveals that these peaks are formed by transitions from the occupied Cu\(^{1+}\) 3\(d_{xz,yz}\) bands to the  Cu\(^{2+}\) 3\textit{d} states with \(xy\) symmetry, see Fig. 11(a). The high intensity and sharpness of the peaks are explained by the low dispersion of the Cu\(^{2+}\) 3\textit{d\(_{xy}\)} states, lying at the Fermi level in the LSDA calculations, and the Cu\(^{1+}\) 3\(d_{xz,yz}\) bands near the \(\Gamma\) points of the Brillouin-zone in Fig. 11(b). The \textit{initial} states with different symmetries, \textit{xz} and \textit{yz}, contribute differently to the optical conductivities \srxx$(\omega)$ and \sryy$(\omega)$, as indicated in Fig. 11. While these transitions have predominantly \textit{d} character,  the contribution of \textit{p} states to the Cu\(^{1+}\) 3\(d_{xz,yz}\) bands is sufficient to provide  a significant transition probability through optical dipole matrix elements.

The above analysis of the interband transitions, which are responsible for the strong anisotropy of the optical properties of {\nco}, is based on the {\it spin-restricted} LSDA calculation which gives a metallic solution with four Cu$^{2+}$ {\dxy}-derived bands crossing the Fermi level (Fig. 8). Although a minute gap of less than 0.1 eV opens in LSDA calculations for {\it spin-spirals} (not shown), it is still much smaller than the experimental gap of 2 eV. The reason for the discrepancy is that the strength of
electronic correlations in the Cu$^{2+}$ $3d$ shell is strongly underestimated
within the LSDA.  When properly accounted for, the on-site Coulomb repulsion
$U_{dd}$ would split the half-filled Cu$^{2+}$ {\dxy} bands into occupied lower and unoccupied upper Hubbard bands and open an insulating gap. Then, the sharp peaks of the calculated optical conductivity, which appear at $\sim$1.5 eV due to the interband transitions involving the Cu\(^{2+}\) {\dxy} \textit{final} states (Fig. 8), would shift to higher photon energies improving the agreement with the measured spectra.

In order to illustrate the effect of the electronic correlations on the
optical conductivity of {\nco} we recalculated its band structure and optical
spectra using the LSDA+$U$ method \cite{AZA91}. The LSDA+$U$ calculations were performed assuming ferromagnetic order of the Cu$^{2+}$ moments. The value of the exchange integral $J$ was fixed to the LSDA value of 1 eV, and the on-site Coulomb repulsion $U_{dd}$ was varied from 2 to 8 eV. Before presenting the results of the LSDA+$U$ calculations it is worth recalling that when the so-called atomic limit is used for the double counting term \cite{cs94} and non-spherical contributions to $U_{dd}$ and $J$ are neglected, the expression for the orbital dependent LSDA+$U$ potential $V_{\sigma i}$, which is to be added to the LSDA potential, becomes particularly simple:
\begin{equation}
V_{\sigma i}=(U_{dd}-J)\left( \frac{1}{2}-n_{\sigma i}\right),
\label{eq:vldau}
\end{equation}
where $n_{\sigma i}$ is the occupation of $i$-th localized orbital with the
spin $\sigma$. One immediately notices that the main effect of LSDA+$U$ is to split occupied ($n_{\sigma i} \approx 1$) and unoccupied ($n_{\sigma i}
\approx 0$) states by shifting the former by $(U_{dd}-J)/2$ downwards and the latter by the same amount upwards with respect to their LSDA energy position.

Since all the Cu$^{2+}$ $d$ states, except for \dxy\ ones, are completely filled already in LSDA, they shift by $(U_{dd}-J)/2$ to lower energies when $U_{dd}$ is applied. The half-filled {\dxy} states contribute substantially to bonding O$p$--Cu $d$ states near the bottom of the valence band because of strong $\sigma$-type hybridization with O $(p_x\pm p_y)\sqrt{2}$ states. In LSDA+$U$ calculations the majority spin {\dxy} state becomes fully occupied and moves by $(U_{dd}-J)/2$ downwards. The occupation of the minority spin {\dxy} state, however, does not go to zero, as one would expect for a formally unoccupied state, but instead remains close to 0.5 due to the strong {\dxy} contribution to the occupied O $p-$Cu $d$ bonding states. According to Eq. (7), the LSDA+$U$ potential acting on the minority spin Cu$^{2+}$ {\dxy} state is much less than $(U_{dd}-J)/2$, and the energy of the bands formed by these states remains close to the LSDA value. In order to avoid such asymmetric splitting of the \dxy\ bands, we performed the LSDA+$U$ calculations with the occupation numbers of the majority and minority spin Cu$^{2+}$ {\dxy} states fixed to 1 and 0, respectively. In this way the corresponding majority and minority spin bands are shifted by $\pm(U_{dd}-J)/2$ with respect to their LSDA position.

The optical conductivities \srxx$(\omega)$ and \sryy$(\omega)$ calculated with $U^{Cu^{2+}}_{dd}$=6 eV acting on the Cu$^{2+}$ $d$ states are presented in  Fig. 12 (b). With this value of $U^{Cu^{2+}}_{dd}$, the unoccupied bands formed by the minority spin Cu$^{2+}$ {\dxy} states lie $\sim$2 eV above the Fermi level and, in contrast to the LSDA result, the LSDA+$U$ band structure is insulating with a gap of about 1.8 eV. The sharp absorption peaks, which are caused by inter-band transitions to the final states of the Cu$^{2+}$ {\dxy} character and appear at $\sim$1.5 eV in the LSDA spectra (Fig. 9), shift to 3 eV in much better agreement with the experiment. The peaks show strong polarization dependence, with the \sryy\ peak being 0.3 eV lower than the \sryy\ one.
When $U^{Cu^{1+}}_{dd}$ of 6 eV is applied also to the $d$ states of Cu$^{1+}$ ions, the bands formed by Cu$^{1+}$ {\dyz} and {\dxz} states, which are the initial states for the interband transitions responsible for the sharp peaks, shift to lower energies. In this case all $n_{\sigma i} \approx 1,$ and LSDA+$U$ mimics to some extent the effect of the so-called self-interaction corrections \cite{PZ81} by shifting all the Cu$^{1+}$ $d$ states downwards.
This brings the conductivity peaks even closer to their experimental
positions. Nevertheless, the magnitude of the calculated optical conductivity
remains lower than in the experiment.
One would also expect some narrowing of the unoccupied Cu$^{2+}$ \dxy\ bands if the LSDA+$U$ calculation were performed for the incommensurate magnetic structure observed experimentally.\cite{Cap05,Cap10} This could result in sharpening of the peaks and an increase of the calculated conductivity.

Our calculations suggest that the origin of the anomalously strong absorption peaks observed at 3.45 and 3.7 eV can be ascribed to the interband Cu\(^{1+}\)
3\(d\) \(\rightarrow\) Cu\(^{2+}\) 3$d$ transitions. The observed shift in the peak positions between two polarizations arises from different contributions
of transitions from the \textit{initial} {\dxz} and {\dyz} states to the optical conductivity, as shown in Fig. 11. The experimentally observed
anisotropy of the background, assigned to the Cu\(^{1+}\) 3\(d \rightarrow\)
4\(p\)\ transitions, is due to the different contributions of transitions to the \textit{final} states with {\px} and {\py} characters to the optical conductivities \srxx$(\omega)$ and \sryy$(\omega)$, as shown in Fig. 10.

\section{Conclusions}

In our spectroscopic ellipsometry study of {\nco}, we observed that the spectra of the dielectric function for light polarized parallel to the $\rm Cu^{1+}$ planes exhibit a strong in-plane anisotropy of the interband excitations. The absorption edge for polarization along the $\rm Cu^{2+}O_2$ chains is formed by a two-peak structure centered at 2.15 and 2.65 eV. This feature is absent in the other polarization and bears striking resemblance to the one observed in the single-valent $\rm Cu^{2+}O_2$ chain compound $\rm LiCuVO_4$.\cite{Mat09} Our findings suggest that these modes are a generic feature of Mott-Hubbard insulators with edge-sharing copper-oxide chains. Based on theoretical considerations, we have also shown that an exciton doublet is expected to emerge as a consequence of the long-range Coulomb interaction at energies $U-V$ and $U-V/2$ upon cooling into a temperature range where substantial spin correlations become established. Identification of this exciton doublet with the experimentally observed two-peak feature allowed us to determine the local Hubbard interaction parameter \(U=3.2\) eV and the long range Coulomb repulsion parameter \(V=1.1\) eV.

The quantitative information about elementary excitations in insulating chain cuprates gained from the present study may deepen our understanding of the electronic structure and phase behavior of doped cuprates as well. In the doped CuO$_2$-planes of the cuprate superconductors the band width is significantly larger, and hence excitons are not expected to appear. Nonetheless, the quantitative description of the effect of the long-range Coulomb interaction obtained here may facilitate realistic calculations of the properties of charge density wave and striped states in doped compounds with one-dimensional \cite{Hor05,Rai08} and two-dimensional electronic structure.

We have also observed strong and sharp absorption bands peaked at 3.45 eV and 3.7 eV  for light polarization along and perpendicular to the $\rm Cu^{2+}O_2$ chains, respectively, dominating the spectra and superimposed on a flat and featureless plateau background above the absorption edge. Based on density functional calculations, we conclude that the major contribution to the background response comes from the intra-atomic Cu\(^{1+}\) 3\(d \rightarrow\) 4\(p\)\ transitions within the {\OCO} dumbbells. The experimentally observed anisotropy of the background is explained due to the different characters of the final 4{\px} and 4{\py} states. Pisarev {\it et al.} have attributed this anomalous absorption feature to exciton formation of the Cu\(^{1+}\) 4$p$ electron - 3$d$ hole pairs.\cite{Pis06} We propose an alternative explanation based on consideration of electron correlations in the LSDA calculations. Our results indicate that the anomalous peak superimposed on the background absorption can be assigned to transitions between bands formed by $\rm Cu^{1+}$3$d_{xz}$/$d_{yz}$ and $\rm Cu^{2+}$3$d_{xy}$ orbitals, which are strongly hybridized with O-$p$ states. In this approach, the observed $\sim$ 0.25 eV shift in the peak positions between two polarizations arises from anisotropic contributions of transitions from the initial $\rm Cu^{1+}$ 3$d_{xz}$ and 3$d_{yz}$ states to the optical conductivity.


\begin{thebibliography}{99}

\bibitem{Mae01} S. Maekawa and T. Tohyama, Rep. Prog. Phys. \textbf{64} 383 (2001).

\bibitem{Miz98} Y. Mizuno, T. Tohyama, S. Maekawa, T. Osafune, N. Motoyama, H. Eisaki, and S. Uchida, Phys. Rev. B \textbf{57}, 5326 (1998).

\bibitem{Mot96} N. Motoyama, H. Eisaki, and S. Uchida, Phys. Rev. Lett. \textbf{76}, 3212 (1996).

\bibitem{Hor05} P. Horsch, M. Sofin, M. Mayr, and M. Jansen, Phys. Rev. Lett. \textbf{94}, 076403 (2005).

\bibitem{Rai08} M. Raichle, M. Reehuis, G. Andr\'{e}, L. Capogna, M. Sofin, M. Jansen, and B. Keimer, Phys. Rev. Lett. \textbf{101}, 047202 (2008).


\bibitem{Bar02} W. Barford, Phys. Rev. B \textbf{65}, 205118 (2002).

\bibitem{Jec03} E. Jeckelmann, Phys. Rev. B \textbf{67}, 075106 (2003).

\bibitem{Geb97} F. Gebhard, K. Bott, M. Scheidler, P. Thomas, and S. W. Koch, Philos. Mag. B \textbf{75}, 47 (1997).

\bibitem{Gal97} F.B. Gallagher and S. Mazumdar, Phys. Rev. B \textbf{56}, 15025 (1997).

\bibitem{May06} M. Mayr, and P. Horsch, Phys. Rev. B \textbf{73}, 195103 (2006).

\bibitem{Zha88} F.C. Zhang and T.M. Rice, Phys. Rev. B \textbf{37}, 3759 (1988).

\bibitem{Mat09} Y. Matiks, P. Horsch, R. K. Kremer, B. Keimer, and A. V. Boris, Phys. Rev. Lett. {\bf 103}, 187401 (2009).

\bibitem{Mal04} A. Maljuk, A. B. Kulakov, M. Sofin, L. Capogna, J. Strempfer, C. T. Lin, M. Jansen, and B. Keimer, J. Cryst. Growth \textbf{263}, 338 (2004).

\bibitem{Cap05} L. Capogna, M. Mayr, P. Horsch, M. Raichle, R. K. Kremer, M. Sofin, A. Maljuk, M. Jansen, and B. Keimer, Phys. Rev. B \textbf{71}, 140402(R) (2005).

\bibitem{Cap10} L. Capogna, M. Reehuis, A. Maljuk, R. K. Kremer, B. Ouladdiaf, M. Jansen, and B. Keimer, Phys. Rev. B \textbf{82}, 014407 (2010).

\bibitem{Lei10} Ph. Leininger, M. Rahlenbeck, M. Raichle, B. Bohnenbuck, A. Maljuk, C. T. Lin, B. Keimer, E. Weschke, E. Schierle, S. Seki, Y. Tokura, and J. W. Freeland, Phys. Rev. B \textbf{81}, 085111 (2010).

\bibitem{Pis06} R. V. Pisarev, A. S. Moskvin, A. M. Kalashnikova, A. A. Bush, and Th. Rasing, Phys. Rev. B \textbf{74}, 132509 (2006).

\bibitem{Elli} J. A. Woollam Co., Inc., Spectroscopic Ellipsometry Data Acquisition and Analysis Software WVASE32 $^{\textregistered}$ [http://www.jawoollam.com].

\bibitem{End05} M. Enderle, C. Mukherjee, B. F\aa k, R. K. Kremer, J.-M. Broto, H. Rosner, S.-L. Drechsler, J. Richter, J. Malek, A. Prokofiev, W. Assmus, S. Pujol, J.-L. Raggazzoni, H. Rakoto, M. Rheinstaedter, and H. M. Ronnow, Europhys. Lett. \textbf{70}, 237 (2005).

\bibitem{End10}M. Enderle, B. F\aa k, H.-J. Mikeska, R. K. Kremer, A. Prokofiev, and W. Assmus, Phys. Rev. Lett. {\bf 104}, 237207 (2010).

\bibitem{Dre11}S.-L. Drechsler, S. Nishimoto, R. O. Kuzian, J. M\'alek, W. E. A. Lorenz, J. Richter, J. van den Brink, M. Schmitt, and H. Rosner, Phys. Rev. Lett. {\bf 106}, 219701 (2011).

\bibitem{End11} M. Enderle, B. F\aa k, H.-J. Mikeska, and R.K. Kremer, Phys. Rev. Lett. {\bf 106}, 219702 (2011).

\bibitem{Koo11} H.J. Koo, C. Lee, M.-H. Whangbo, G. J. McIntyre and R. K. Kremer, Inorg. Chem. {\bf 50}, 3582 (2011).


\bibitem{Kir89} J. Kircher, M. Alouani, M. Garriga, P. Murugaraj, J. Maier, C. Thomsen, M. Cardona, O. K. Andersen, and O. Jepsen, Phys. Rev. B {\bf 40}, 7368 (1989).

\bibitem{Kel89} M. K. Kelly, P. Barboux, J.-M. Tarascon, and D. E. Aspnes, Phys. Rev. B {\bf 40}, 6797 (1989).

\bibitem{Mas04} T. Masuda, A. Zheludev, B. Roessli, A. Bush, M. Markina and A. Vasiliev, Phys. Rev. B {\bf 72}, 014405 (2005).

\bibitem{Zvy02}S. Zvyagin, G. Cao, Y. Xin, S. McCall, T. Caldwell, W. Moulton, L.-C. Brunel, A. Angerhofer, and J. E. Crow, Phys. Rev. B {\bf 66}, 064424 (2002).

\bibitem{And75} O. K. Andersen, Phys. Rev. B \textbf{12}, 3060 (1975).

\bibitem{AZA91} V.I. Anisimov, J. Zaanen, and O. K. Andersen, Phys. Rev. B \textbf{44}, 943 (1991).

\bibitem{cs94} M. T. Czy\.zyk and G. A. Sawatzky, Phys. Rev. B \textbf{49}, 14211 (1994).

\bibitem{PZ81} J.P. Perdew and A. Zunger, Phys. Rev. B \textbf{23}, 5048 (1981).

\end{thebibliography}
\end{document}